\documentclass[a4paper]{jpconf}
\usepackage{amsmath}
\usepackage{algorithm}
\usepackage{algorithmic}
\usepackage{graphicx}
\begin{document}
\title{Finding a Hadamard matrix by simulated annealing of spin-vectors}

\author{Andriyan Bayu Suksmono}

\address{School of Electrical Engineering and Informatics, Institut Teknologi Bandung, Jl. Ganesha No.10, Bandung, Indonesia
}

\ead{suksmono@stei.itb.ac.id, suksmono@yahoo.com}

\begin{abstract}
Reformulation of a combinatorial problem into optimization of a statistical-mechanics system, enables finding a better solution using heuristics derived from a physical process, such as by the SA (Simulated Annealing). In this paper, we present a Hadamard matrix (H-matrix) searching method based on the SA on an Ising model. By equivalence, an H-matrix can be converted into an SH (Seminormalized Hadamard) matrix; whose first columns are unity vector and the rest ones are vectors with equal number of $-1$ and $+1$ called SH-vectors. We define SH spin-vectors to represent the SH vectors, which play a similar role to the spins on an Ising model. The topology of the lattice is generalized into a graph, whose edges represent orthogonality relationship among the SH spin-vectors. Started from a randomly generated quasi H-matrix $Q$, which is a matrix similar to the SH-matrix without imposing orthogonality, we perform the SA. The transitions of $Q$ are conducted by random exchange of $\{+,-\}$ spin-pair within the SH-spin vectors which follow the Metropolis update rule. Upon transition toward zero-energy, the $Q$-matrix is evolved following a Markov chain toward an orthogonal matrix, at which point the H-matrix is said to be found. We demonstrate the capability of the proposed method to find some low order H-matrices, including the ones that cannot trivially be constructed by the Sylvester method.
\end{abstract}

\section{Introduction}
A Hadamard matrix (H-matrix) is an orthogonal binary matrix whose entries are -1 or +1. The matrix has been studied by Sylvester \cite{Sylvester1867} and was later realized by Jacques Hadamard during his investigation on maximal determinant problem \cite{Hadamard1893}. Unique properties of the H-matrix attracts many mathematicians, scientists, and engineers to study it more intensively. The orthogonality and binariness of the matrix enable various kind of practical applications. In thedigital communications area, the Hadamard-Walsh code has been used in CDMA (Code Division Multiple Access) systems, such as the IS-95 or CDMA2000 \cite{Garg2007}, \cite{Seberry2005}. The H-matrices has also been used to construct an error-correction code, which capable to correct large errors suffered by the communicated data. Due to this capability, in 1971 NASA employed the Hadamard code for the Mariner 9 space probe to send digital images of the Planet Mars across the deep space toward a receiver on the Earth. 

One of the most important problems in the study of the H-matrix is on its existence. It is well known that $2^k$-order H-matrices for any positive integer $k$ are exist, in which the Sylvester's method can be used for the construction. Various H-matrix construction methods for order other than $2^k$ have also been proposed; such as Paley's finite field \cite{Paley1933}, Dade-Goldberg's group permutation \cite{DadeGoldberg1959}, Williamson method \cite{Williamson1944}, Bush's finite projective plane \cite{Bush1971a}, \cite{Bush1971b}, and Wallis's orthogonal design \cite{Wallis1976}. However, there is no general method for constructing the $4k$-order H-matrix. The Hadamard Matrix Conjectures states that there is a $4k$-order H-matrix for every positive integer $k$. At present, the largest known H-matrix for a general $4k$-order is 428, which was discovered by Karaghani and Tayfeh-Rezaie \cite{Karaghani2005}.

In principle, provided that the Hadamard-conjecture is true, one can find a $4k$-order H-matrix by performing orthogonality tests to all of the $4k$-order binary matrices. The total number $N_B$ of such $4k$ order binary matrix is 

\begin{equation}
\label{EQ_1}
  N_B\left(4k\right) = 2^{16k^2}
\end{equation}

Therefore, in the worst case condition, the search space is of exponential order. On the other hand, verifying that a binary $4k$-order matrix is Hadamard (or not), one needs only a matrix multiplication whose complexity is of polynomial order. Consequently, finding a Hadamard matrix is an NP (Nondeterministic Polynomial) problem. 

The usage of SA (Simulated Annealing) to solve hard optimization problems have been initiated by Kirkpatrick \textit{et al.} \cite{Kirkpatrick1983} and Cerny \cite{Cerny1983}, who realized the similarity and a deep-connection between combinatorial problems and the statistical mechanics.  The optimization method proposed by these authors employs Monte-Carlo algorithm with Metropolis update-rule \cite{Metropolis1953}. For other practical applications, the SA has also been used in image restorations; both to restore the real-valued images \cite{Geman1984} and the complex-valued ones \cite{Suksmono2002}. 

In this paper, we present a method to find a Hadamard matrix by using SA. The distinctive feature in our method is; instead of using binary values on an Ising lattice, we consider a special binary vectors, or spin-vectors, which have balanced number of -1 and +1 spins. By started from a system of a set of randomly generated $(4k-1)$ spin-vectors, the SA seek for a configuration with global minimum energy of the spin-vector system. When the ground-state is achieved, the set of vectors become orthogonal and a $4k$-order H-matrix can be constructed. At this condition, we say that the H-matrix has been found.

The rest of the paper is organized as follows. In Section 2, we briefly discuss the basic properties of an H-matrix, SH vectors, and define SH-spin vectors employed throughout the paper. The proposed method is formulated in Section 3, where an SA algorithm for the SH spin-vectors case is sketched. Search examples of a few low order H-matrix, including the ones that cannot be constructed by the Sylvester method, are presented in Section 4. In the final part, the paper will be concluded in Section 5.

\section{The Hadamard Matrix and Seminormalized Hadamard Spin-Vectors}

\subsection{SH Matrix, SH Vectors, and Quasi H-Matrix}

An $m$-order H-matrix is an $m \times m$ orthogonal matrix whose entries are either -1 or 1. We will be writing explicitly the sign on the entries, so that the number 1 is written as +1 for clarity of explanations. We also assume that $m=4k$, with $k$ a positive integer, throughout the paper. The orthogonality of the H-matrix means that the following relationship is held

\begin{equation}
\label{EQ_2}
 H^TH= m I
\end{equation}
where I is an $m \times m$ identity matrix. For an H-matrix, reordering (exchanging or permuting) the rows (or columns), transposition, and/or negation (multiplication by -1) of the rows (or columns) yields another H-matrix; i.e., these operations do not change the status of the H-matrix. The H-matrices obtained by such operations are called equivalent H-matrices, and the operations will be called as equivalent operations.

By the equivalent operations, an H-matrix can be normalized or seminormalized. An H-matrix is normalized, shortly called an NH-matrix, if the entries of both the first column and the first row are +1. Whereas an H-matrix is seminormalized, written as an SH-matrix, if the entries of either its first column or first rows, are +1. For consistency, we will choose the SH-matrix as an H-matrix whose first column’s entries are +1 or a unity vector. Furthermore, we will write +1 entries as “+” and the -1 as “-“ for conciseness. The followings are equivalent operations to an H-matrix that yields an SH-matrix and an NH-matrix.

\begin{equation}
\label{EQ_X}
H_1=
\begin{pmatrix}
  + & + & - & -\\
  + & - & - & +\\
  - & - & - & -\\
  + & - & + & -
\end{pmatrix} 
\to
H_2=
\begin{pmatrix}
  + & + & - & -\\
  + & - & - & +\\
  + & + & + & +\\
  + & - & + & -
\end{pmatrix}
\to
H_3=
\begin{pmatrix}
  + & + & + & +\\
  + & - & - & +\\
  + & + & - & -\\
  + & - & + & -
\end{pmatrix}
\nonumber
\end{equation}
In this example, $H_1$, $H_2$, and $H_3$ are equivalent H-matrices. In particular, $H_2$ is an SH-matrix which is obtained by negating the third row of $H_1$, whereas $H_3$ is an NH-matrix which is obtained by exchanging the first row with the third one of $H_2$. 

For an SH-matrix, since the first column is a unity vector, the orthogonality condition in Eq. (\ref{EQ_2}) implies that the other column vectors must have a balanced number of -1 and +1; i.e. they are $4k$ order SH (Seminormalized Hadamard) vectors whose $2k$ entries are +1 and the rest of $2k$ number of entries are -1.  Based on basic counting, the number of $4k$-order SH vector $N_V$ is equal to the number of ways to arrange $2k$ number of objects; i.e. -1, into $4k$ positions. Then, we obtain the following result

\begin{equation}
\label{EQ_3}
N_V\left(4k\right) = C(4k,2k) =\frac{4k!}{(2k)!(2k)!}
\end{equation}

In the previous example, the $2^{nd}$, $3^{rd}$, and $4^{th}$ columns of $H_2$ and $H_3$ are SH vectors. The complete list of 4-order SH vectors, which according to Eq.(\ref{EQ_3}) will consist of $N_V=6$ distinct vectors, are as follows:

$\vec{v}_1=\left( \begin{array}{cccc} +&+&-&-\end{array}\right)^T$, $\vec{v}_2=\left( \begin{array}{cccc} -&-&+&+\end{array}\right)^T$,$\vec{v}_3=\left( \begin{array}{cccc} +&-&+&-\end{array}\right)^T$, 
 
$\vec{v}_4=\left( \begin{array}{cccc} +&-&-&+\end{array}\right)^T$, $\vec{v}_5=\left( \begin{array}{cccc} -&+&+&-\end{array}\right)^T$, $\vec{v}_6=\left( \begin{array}{cccc} -&+&-&+\end{array}\right)^T$.

A pair of (randomly generated) SH-vectors are generally not orthogonal. In analogy to the SH-matrix, we define a square binary matrix $Q$ whose first column is a unity vector and the rest ones are SH-vectors, which is called a quasi H-matrix or the Q-matrix. In general, $Q$ is non-orthogonal and at a very special condition when it is orthogonal, $Q$ becomes an H-matrix. It can be showed that the number $N_{QU}(4k)$ of unique $4k$-order quasi-H-Matrix can be approximated as follows

\begin{equation}
\label{EQ_4}
N_{QU}(4k) \approx \left( \frac{2^{4k}}{8k^{3/2}}\right)^{4k}
\end{equation}
Compared to the number of binary matrix of the corresponding order given by Eq. (\ref{EQ_2}), the number of unique $Q$ is relatively smaller.

\subsection{SH-Spin Vectors}
Usually, an Ising lattice consists of two kinds of spin, i.e., the spin-up +1 and spin-down -1. We need to define a more general spin notation that consists of a group of $4k$ number of spins and treating it as a single entity. The purpose to define such an object is to represent the SH-vectors as an element (entry) of H-matrix on an Ising model. The main idea of finding H-matrix by the SA is evolving a binary non-orthogonal Q-matrix through a Markov-chain by randomly changing the vectors, from one SH-vector to another SH-vector, followed by detection of the orthogonality of the matrix. When the orthogonality condition is achieved, the system is said to reach a ground state. Since a particular SH vector should only change into another SH-vector, the single-spin-flip at a time suggested in \cite{Newman1999} is not sufficient; instead, a pair of $-$ and $+$ spins in the SH spin-vector should be flipped or exchanged. Therefore, the SH-spin vectors are objects on the Ising lattice that posses the following properties:
\begin{itemize}
\item It is a vector that consist of $2k$ number of spin-up (+1), and $2k$ number of spin down (-1)
\item The change or transition of a spin-vector should involve a pair of spin-up and spin-down, so that the SH-vector property is preserved.
\end{itemize}
For illustration, the following transition of exchanging the second spin with the third one $\left( \begin{array}{cccc} +&+&-&-\end{array}\right)^T \to$ $\left( \begin{array}{cccc} +&-&+&-\end{array}\right)^T$ is allowed, however, the following transition $\left( \begin{array}{cccc} +&+&-&-\end{array}\right)^T \to$ $\left( \begin{array}{cccc} +&+&+&-\end{array}\right)^T$ of flipping only the third spin is forbidden since the result is not an SH-vectors, i.e., the numbers of -1 and +1 are not balanced anymore.

\section{Simulated Annealing of SH Spin-Vectors }

\subsection{Scalar Spin with Single-Spin Flip}
The Ising model is basically a statistical mechanics model of ferromagnetism, which not only exhibits phase transition phenomena (in two-dimensional, Peierls \cite{Peierls1936}), but also are exactly solvable for some particular cases \cite{Onsager1944}, \cite{Baxter1982}. Recently, Cuevas \textit{et al.} proposed the Ising model as a universal model that capable to capture all aspects of classical spin physics \cite{Cuevas2016}.

In an Ising model the spins are normally arranged on a regular grid (lattice) and each of the spins interact with its neighbors, which is showed in Fig.1. (a). 
\begin{figure}
 \label{FIG_1}
 \centering
 \includegraphics[scale=1.0]{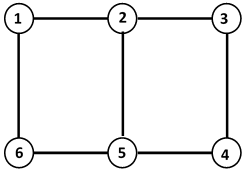}
 \includegraphics[scale=1.4]{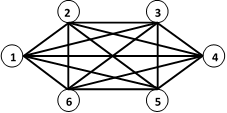}
 
  \begin{tabular}{cccccccccccccccc}
  \\
  (a) & & & & & & & & & & & & & & (b) \\
 \end{tabular}

 \caption{Graphical illustration of the Ising model and connection among the spins: (a) nearest neighbor connections (b) fully connected spins}
\end{figure}
The energy of the Ising model for a particular spin configuration $s$ is given by

\begin{equation}
\label{EQ_5}
 E(s)=-\sum_{<i,j>} J_{ij}s_i s_j - \mu \sum_{j} h_j s_j
\end{equation}
where $\left<i,j\right>$ denotes the summation over the products of different $s_i$ and $s_j$, $J_{ij}$ is the interaction strength or coupling between spins at site-$i$ and the one at site-$j$, while $h_j$ is the interaction strength between external magnetic field $\mu$ with a spin $s_j$. When considering nearest neighbor interaction, the non-zero value of $J_{ij}$ is generally applied only to the directly or neighboring connected pairs, so that in Fig.1 (a), for examples, the $J_{1,2}$ and $J_{1,6}$ are non-zeros, while $J_{1,3}=0$. In this paper, we consider a more general arrangement of the spins, in term of configuration and interaction. The spins are represented as vertices of a graph, whereas the interaction is represented by edges of the graph. A fully connected spins will be represented by a complete graph, such as shown in Fig.1 (b). 

The configurations of the Ising model is initialized by a randomly generated Q-matrix $Q$. Since the SA minimization should evolve $Q$ from a high energy state down to the ground-state, where $Q$ becomes orthogonal, the ground-state can be considered as a condition of the system's energy of becoming zero, whereas non-orthogonality of $Q$ should be associated with a positive (higher) energy state. Therefore, in a single-flip-spin scheme, we may define the energy of an $m=4k$ order $Q$, which indicates the deviation of $Q$ from an orthogonal matrix as follows. First, we define an indicator matrix $D\left(Q\right)=Q^TQ$ whose entry is denoted by $d_{ij}\left(Q\right)$. Then, the energy is given by

\begin{equation}
\label{EQ_6}
 E(Q)= \sum_{i} \sum_{j} \left| d_{ij}(Q) \right| - m^2
\end{equation}
where $\left| \cdot \right|$ denotes the absolute value. If $Q$ is orthogonal, then the first term on the right hand side expresses the sum of entries of a diagonal matrix, and therefore $E(Q)=0$ as desired.

\begin{figure}
 \label{FIG_2}
 \centering
 \includegraphics[scale=0.4]{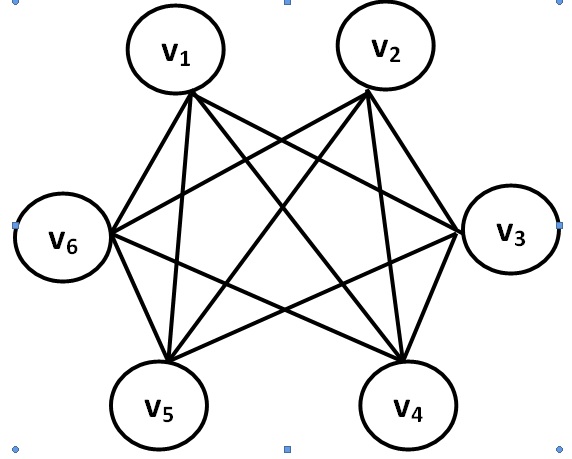}
 \caption{Orthogonality graph of the 4-order SH vectors}
\end{figure}

\subsection{SH-Vector Spin with Pair-of-Spin Flip}
Alternatively, we can consider an orthogonality graph of $4k$-order SH vectors, such as displayed in Fig.2 for a 4-order SH vectors. Each of the $(4k-1)$ vertices of the complete subgraph can be used to construct an SH-matrix $H$, such as $\left\{ v_1, v_3, v_4 \right\}$, by inserting a unity vector as the first column whereas this set of SH vectors belong to the other columns. In the SA, we use a set of randomly generated $(4k-1)$ SH-vectors, which are then connected as a $(4k-1)$ complete graph to construct an SH-Ising vectors lattice, followed by evolving them by using SA algorithm so that they subsequently become orthogonal to each other. The energy of such a configuration can be written as:

\begin{equation}
\label{EQ_7}
 E(\vec{v}) = \sum_{i\neq j} \left| \vec{v}_i \cdot \vec{v}_j \right| + \sum_{i}\left| \vec{1} \cdot \vec{v_i}\right|
\end{equation}
The second term in Eq.(\ref{EQ_7}) ensures that the evolution maintain the vectors $v_i$ of being an SH-vectors. Since we restrict $v_i$ to only SH-vectors, the spin-flip on the SH-spin vector should not change the number of +1 and -1 entries. It can be achieved only if we flip them in pairs, or exchange the position of the pair.

\begin{figure}
 \label{FIG_3}
 \centering
 \includegraphics[scale=0.7]{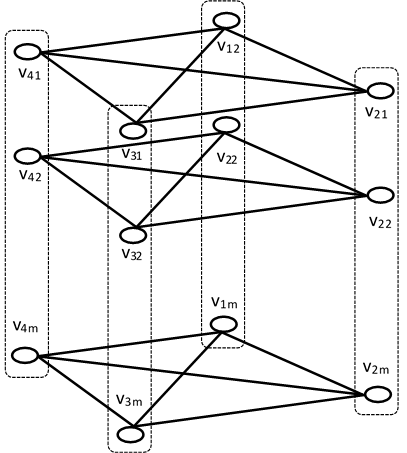}
 \includegraphics[scale=0.7]{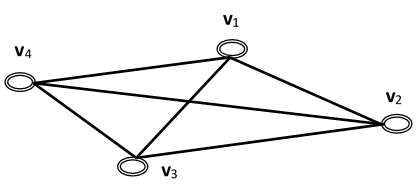}
   \begin{tabular}{cccccccccccccccc}
  \\
  (a) & & & & & & & & & & & & & & (b) \\
 \end{tabular}

 \caption{Illustration of connections among the SH-spin vector on an Ising model: (a)SH-spin vector connections among the individual spin,(b) each group of spins are represented as a single SH-spin vectors drawn as double-lined circles, then they are connected as a complete graph.
}
\end{figure}
Figure 3 displays the connections among SH-spin vectors. In (a), the connections is drawn among the individual spins, whereas in (b), they are represented by connection among SH-spin vectors. The pseudocode listed in of Algorithm.1 illustrates the proposed SH-spin vectors SA method.

%==== ALGORITHM GOES HERE =====
\begin{algorithm}
  \label{ALG_SA}
  \caption{H-Matrix Search by Simulated Annealing}
  \begin{algorithmic}[1]
	\STATE \textbf{Input:} Order of SH matrix $4k$, annealing schedule $p_T(t)$, $MaxIter$  	
	\STATE \textbf{Output:} A $4k$-order SH-matrix
  	\STATE Randomly generate $4k$-order Q-matrix $Q$
  	\STATE $t \leftarrow 0$
	\WHILE{$E(Q)>0$ or $t < MaxIter$} 
	  \STATE Copy $Q$ into a template: $Q_{t} \leftarrow Q$
	  \STATE Randomly select a pair of  $\{+1,-1\}$ at a randomly selected column of $Q_t$, then flip (or exchange) the pair.
	  \IF{$E(Q_t)<E(Q)$ or $random >p_T(t)$}
		\STATE $Q \leftarrow Q_t$
	  \ENDIF
	  \STATE $t \leftarrow t+1$
	\ENDWHILE	 	 
	\end{algorithmic}
\end{algorithm}

In the SA algorithm, we have used a scheduled annealing. Initially, the system is set to a total randomness by setting the value of transition threshold of the Metropolis to $0.5$. Then, the threshold is increased subsequently and reaching $1.0$ at the end of the iteration. A detailed example of the schedule is described in the following Section.

\section{Experiments and Analysis}
In the experiment, we set the algorithm to find an H-matrix of order 12, which is the lowest order of an H-matrix that cannot be constructed by the Sylvester method. The SA schedule is displayed in Fig.4 (a) as a threshold function, whereas evolution of the energy curve of $Q$ during the SA is displayed in Fig.4 (b). We observe fluctuation in the energy curve, indicating the Metropolis update scheme works as expected. We also find the trends of decreasing energy in the Fig.4 (b), although it fluctuates over time; which is large in the beginning and suppressed at latter iterations. The system finally reach the ground-state at iteration number around $70,000$ with $E(Q)=0$, indicating that an H-matrix (or SH-matrix) has been found.

\begin{figure}
 \label{FIG_4}
 \centering
 \includegraphics[scale=0.35]{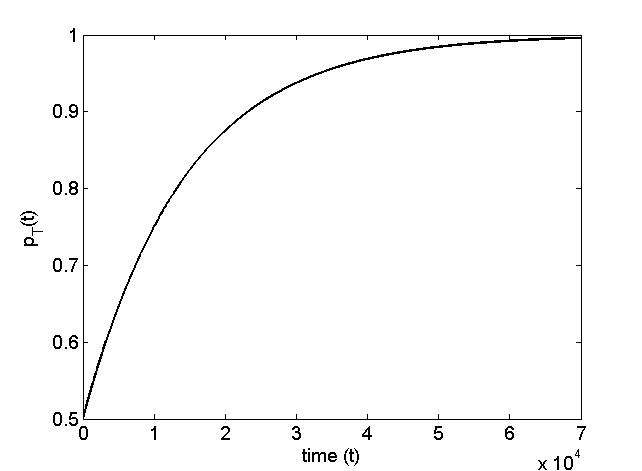}
 \includegraphics[scale=0.35]{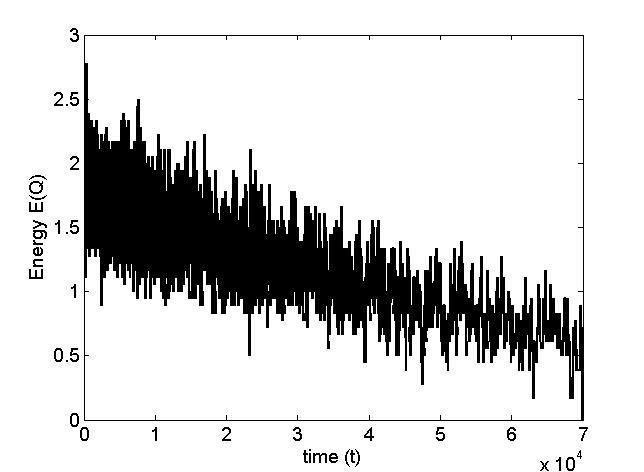}
   \begin{tabular}{cccccccccccccccc}
  \\
  (a) & & & & & & & & & & & & & & (b) \\
 \end{tabular}

 \caption{Execution of the algorithm. Curve of (a) Annealing schedule expressed in $p_T(t)$ and (b) Evolution of the energy (normalized by $m^2$)}
\end{figure}

The initial 12-order randomly generated $Q$ is displayed in Fig.5 (a), where black squares indicate $–1$ entries, whereas the white ones correspond to $+1$. The orthogonality level among the SH vectors of $Q$ are displayed in Fig.5 (b), with black indicates orthogonal, whereas other color/gray level are non-orthogonal vectors. After reaching the ground-state, the H-matrix is found and it is displayed in Fig.6 (a) whose orthogonality is indicated by the diagonal form of indicator matrix in Fig.6 (b).

\begin{figure}
 \label{FIG_5}
 \centering
 \includegraphics[scale=0.3]{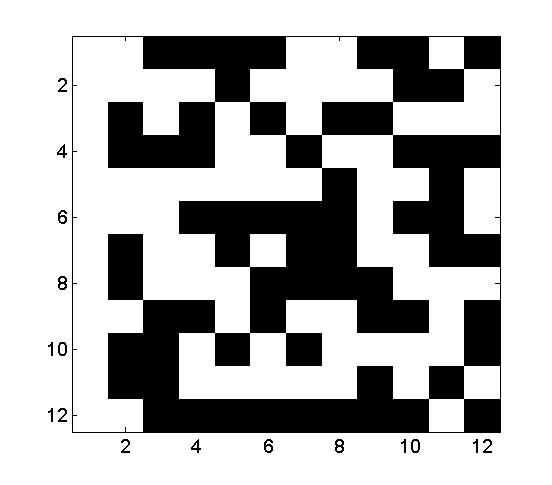}
 \includegraphics[scale=0.3]{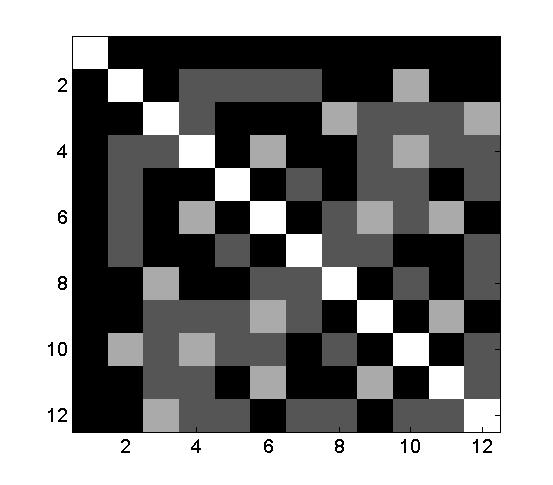}
   \begin{tabular}{cccccccccccccccc}
  \\
  (a) & & & & & & & & & & & & & & (b) \\
 \end{tabular}

 \caption{ Initial condition:(a) randomly generated Q-matrix, (b) Orthogonality indicator matrix $D=Q^TQ$ }
\end{figure}
%
%%  table-1 goes here
\begin{table}
 \caption{\label{Table1}Performance of the proposed method}
  \begin{center}
   \begin{tabular}{ccccc}
   \br
$k$ & Order & NIter & Log(NIter) & MaxIter\\
\mr
1 & 4 & 9 & 0.95 & 16,000 \\
2 & 8 & 4,637 & 3.67 & 64,000 \\
3 & 12 & 69,900 & 4.84 & 144,000 \\
4 & 16 & 1,551,886 & 6.19 & 2,560,000 \\
5 & 20 & 29,548,458 & 7.47 & 40,000,000 \\

   \br
  \end{tabular}
 \end{center}
\end{table}

Table 1 shows performance of the algorithm for the first 5 lowest orders of H-matrices, including the 12-order and 20-order that cannot be constructed by the Sylvester method. The table shows that iteration time grows non-linearly, whereas taking the logarithm indicates the close relationship between $k=m/4$ with it, so that we can expect the grow is like  $O(e^{c\cdot k})$ for an arbitrary constant $c$. The column MaxIter shows the number of maximum iteration set in the program for a given instance.
\begin{figure}
 \label{FIG_6}
 \centering
 \includegraphics[scale=0.3]{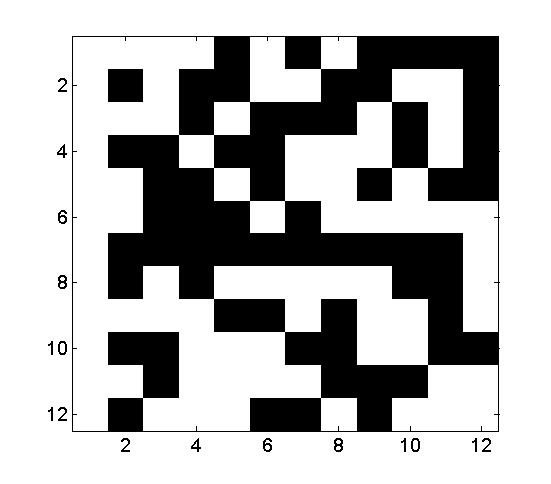}
 \includegraphics[scale=0.3]{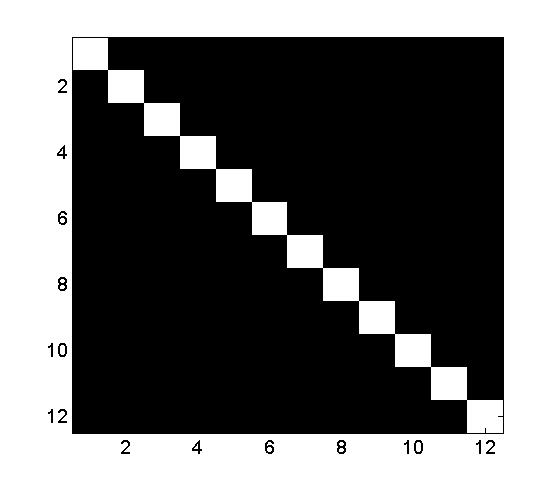}
   \begin{tabular}{cccccccccccccccc}
  \\
  (a) & & & & & & & & & & & & & & (b) \\
 \end{tabular}

 \caption{Result of the algorithm for 12-order H-matrix search: (a)Found H-matrix (b) Indicator matrix shows that $Q$ is now orthogonal}
\end{figure}
\begin{figure}
 \label{FIG_7}
 \centering
 \includegraphics[scale=0.6]{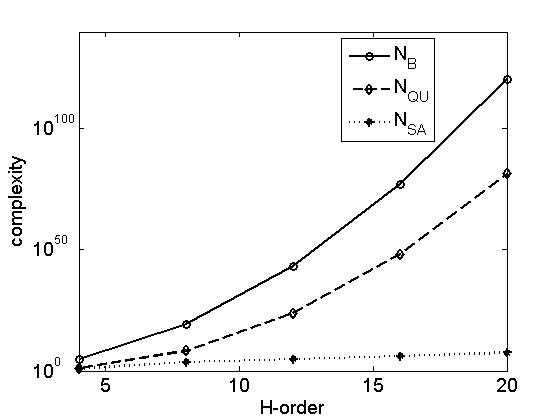}
 \caption{Comparison of $N_B$, $N_{QU}$, and the number of SA iterations to find the H-matrix $N_{SA}$}
\end{figure}

We compare the iteration numbers to find the H-matrix (actually an SH-matrix) $N_{SA}$ with the number of binary matrix $N_B(4k)$  given by Eq.(1) and the number of unique quasi H-matrix $N_{QU}(4k)$ given by Eq.(\ref{EQ_4}), which is displayed in Fig.7. The curves show that iteration number of finding a H-matrix grows much slowly than both of the number of the binary matrix and unique Q-matrix. Therefore, performing SA is much more efficient than exhaustively checking of either all $4k$-order binary matrix or $4k$-order Q-matrix. Nevertheless, the complexity of finding an H-matrix using SA seems to grow exponentially by $O(e^{c\cdot k})$, which makes the algorithm not effective anymore for finding a high order H-matrix. Implementing the SA code on parallel or cluster might help finding a few more high order matrices, but to find the lowest order unknown H-matrix, i.e. 668, we have to consider a quantum computer.  The adiabatic-quantum computer, like the D-Wave system \cite{Johnson2011}, with sufficient connections and number of qubits is prospective for finding such a high order H-matrix.

\section{Conclusions and Further Directions}
We have presented a searching method for finding a Hadamard matrix using SA algorithm. Replacing the individual spins with an SH-spin vector is found effective in finding the SH matrix, which is applied successfully to find the first 5 lowest order H-matrices. Although the method is applicable to a general order of H-matrix, the execution time grows nonlinearly. To find high order H-matrix, we may consider a computer with multiprocessor or a cluster computer, since the SA itself is in principle parallelizable. It is also worth to consider QSA (Quantum SA) algorithm to do the search and use adiabatic quantum computing to find such a high order matrix in the future.

%%%%%%%%%%%%%%%%%%%%%%%%%%%%%%%%%%%%%%%%%%%%%%

\end{document}